# Hybrid Graphene-Gyroelectric Structures: A Novel Platform for THz Applications

## Mohammad Bagher Heydari [1,*]

[1,*] School of Electrical Engineering, Iran University of Science and Technology (IUST), Tehran, Iran

[*] Corresponding author: mo_heydari@alumni.iust.ac.ir

**Abstract:** This paper investigates tunable magneto-plasmons in graphene-based structures combined with gyro-electric layers. In the general waveguide, each graphene sheet has been sandwiched between two different gyro-electric layers. The whole structure has been exposed in the presence of a magnetic bias. An accurate analytical model based on all propagating modes has been proposed and led to closed-form complicated relations. As a special case of the multi-layer waveguide, a hybrid graphene-gyroelectric structure is studied in this paper. It has been illustrated that the propagating features of the exemplary waveguide can be tuned by changing the external magnetic bias and the chemical potential of graphene, which makes it a new tunable platform for the design of non-reciprocal plasmonic devices in the THz region.

## 1. Introduction

Nowadays, graphene is known as a fascinating two-dimensional material for the fabrication and development of innovative photonic components, due to its unique properties [1-4]. Among these features, graphene conductivity, a tunable parameter via electrostatic or magnetostatic bias, has created a new emerging science known as "Graphene Plasmonics". Based on the propagating properties of Surface Plasmon Polaritons (SPPs) on graphene, various plasmonic components have been designed and investigated in the THz frequencies such as sensors [5, 6], couplers [7-9], filters [10-12], resonators [13-15], and circulators [16-19]. Among these devices, graphene-based waveguides play a remarkable role in graphene plasmonics, which are divided into various platforms such as planar [20-28], cylindrical [29-33], and elliptical structures [34-37]. It should be noted that graphene plasmonics is mostly used in the mid-infrared region whereas metal-based plasmonics is a promising candidate in near-infrared frequencies for various applications [38-42].

An effective method to enhance the performance of graphene-based components is by integrating graphene with new smart materials such as chiral materials [43-50], and non-linear materials [51-61]. For instance, the authors in [44] have introduced a new platform for THz sensing applications by integrating graphene sheets with chiral materials. The integration of graphene with new materials will lead to new plasmonic wave phenomena and effects and also will give more degrees of freedom to the designer for the tune of the propagation features of SPPs.

In this paper, we aim to study the plasmonic features of hybrid graphene-gyroelectric structures by using a new analytical method. To the author's knowledge, no published work has been reported for the analytical model of gyro-electric graphene-based multi-layer waveguides. In the general structure, each graphene sheet has been sandwiched between two different gyro-electric layers, with an anisotropic permittivity tensor magnetized by a magnetic bias. This general configuration can be a promising platform for the design of non-reciprocal THz devices because its properties can be altered via magnetic bias and chemical potential. The proposed analytical model considers all propagating modes inside the structure. It gives closed-form relations for the field contributions of all SPP waves. Hence, our analytical model is accurate. Moreover, it can calculate the modal properties of the gyro-electric graphene-based structure faster than any software. Because all electromagnetic software utilizes numerical methods, which need the discretization of the problem.

The remainder of the paper is organized as follows. In section 2, the general structure and its theoretical model will be studied. Then, to show the validity of the proposed model, our analytical results will be compared with the results of [62, 63] in section 3. A novel gyro-electric graphene-based waveguide, as a special case of the general structure, will be introduced and investigated in section 4. Finally, section 5 concludes the article.

## 2. The Proposed General Structure and its Analytical Model

Fig. 1 illustrates the cross-section of the proposed general structure. In this configuration, each graphene sheet has been sandwiched between two different magnetized gyro-electric layers with the following permittivity tensor ($\bar{\bar{\varepsilon}}_N$):

$$\bar{\bar{\varepsilon}}_N = \varepsilon_0 \begin{pmatrix} \varepsilon_N & j\varepsilon_{a,N} & 0 \\ -j\varepsilon_{a,N} & \varepsilon_N & 0 \\ 0 & 0 & \varepsilon_{\parallel,N} \end{pmatrix} \tag{1}$$

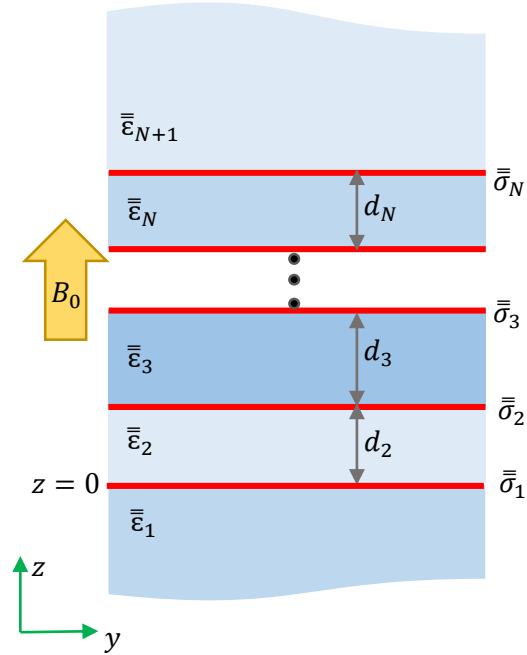

**Fig. 1.** Cross-section of a general gyro-electric multi-layer structure containing graphene sheets. The external magnetic bias is applied in the z-direction.

The elements of permittivity tensor (relation (1)) are considered as follows:

$$\varepsilon_N = \varepsilon_{\infty,N}\left(1 - \frac{\omega_{p,N}^2(\omega + j\upsilon_N)}{\omega\left[(\omega + j\upsilon_N)^2 - \omega_c^2\right]}\right) \tag{2}$$

$$\varepsilon_{a,N} = \varepsilon_{\infty,N}\left(\frac{\omega_{p,N}^2 \omega_c}{\omega\left[(\omega + j\upsilon_N)^2 - \omega_c^2\right]}\right) \tag{3}$$

$$\varepsilon_{\parallel,N} = \varepsilon_{\infty,N}\left(1 - \frac{\omega_{p,N}^2}{\omega(\omega + j\upsilon_N)}\right) \tag{4}$$

In (2)-(4), $\nu_N, \varepsilon_{\infty,N}$ are the effective collision rate and the background permittivity, respectively. Furthermore, the plasma and the cyclotron frequencies are defined as follows [64]:

$$\omega_{p,N} = \sqrt{\frac{n_s e^2}{\varepsilon_0 \varepsilon_{\infty,N} m^*}} \tag{5}$$

$$\omega_c = \frac{e B_0}{m^*} \tag{6}$$

In (5)-(6), $e, m^*$ and $n_s$ are the charge, effective mass, and the density of the carriers, respectively. The conductivity of magnetically biased graphene sheet in the $N$-th layer is expressed as follows [65]:

$$\bar{\bar{\sigma}}_N\left(\omega, \mu_c(E_0), \Gamma, T, B_0\right) = \begin{pmatrix} \sigma_{O,N} & \sigma_{H,N} \\ -\sigma_{H,N} & \sigma_{O,N} \end{pmatrix} \tag{7}$$

In this relation, $\omega$ is the radian frequency, $\Gamma$ is the phenomenological electron scattering rate ($\Gamma = 1/\tau$, where $\tau$ is the relaxation time), $T$ is the temperature, $\mu_c$ is the chemical potential that can be altered by chemical doping or electrostatic bias $E_0$, and $B_0$ is the applied magnetostatic bias field. Besides, $\sigma_{O,N}, \sigma_{H,N}$ are the direct and indirect (Hall) conductivities of the graphene, which are obtained by using Kubo's relations reported in [65].

Consider Maxwell's equations in the frequency domain (suppose $e^{i\omega t}$):

$$\nabla \times \mathbf{E} = -j\omega \mu_N \mathbf{H} \tag{8}$$

$$\nabla \times \mathbf{H} = j\omega \bar{\bar{\varepsilon}}_N \cdot \mathbf{E} \tag{9}$$

The z-components of the electromagnetic fields inside the gyro-electric media can be obtained as:

$$\left(\nabla_\perp^2 + \frac{\varepsilon_{\|,N}}{\varepsilon_N}\frac{\partial^2}{\partial z^2} + (k_0^2 \varepsilon_{\|,N}\mu_N)\right) E_z + \frac{k_0 \mu_N \varepsilon_{a,N}}{\varepsilon_N}\frac{\partial H_z}{\partial z} = 0 \tag{10}$$

$$\left(\nabla_\perp^2 + \frac{\partial^2}{\partial z^2} + (k_0^2 \varepsilon_{\perp,N}\mu_N)\right) H_z - \frac{k_0 \varepsilon_{\|,N}\varepsilon_{a,N}}{\varepsilon_N}\frac{\partial E_z}{\partial z} = 0 \tag{11}$$

where

$$\nabla_\perp^2 = \frac{1}{r}\frac{\partial}{\partial r} r \frac{\partial}{\partial r} + \frac{1}{r^2}\frac{\partial}{\partial^2 \varphi} \tag{12}$$

$$\varepsilon_{\perp,N} = \varepsilon_N - \frac{\varepsilon_{a,N}^2}{\varepsilon_N} \tag{13}$$

By considering the radial modes inside the gyro-electric layer,

$$H_{z,N}(r,\varphi,z) = \int_{-\infty}^{+\infty} \sum_{m=-\infty}^{\infty} H_{z,N}^m(z) \exp(-jm\varphi) J_m(\beta r) d\beta \tag{14}$$

$$E_{z,N}(r,\varphi,z) = \int_{-\infty}^{+\infty} \sum_{m=-\infty}^{\infty} E_{z,N}^m(z) \exp(-jm\varphi) J_m(\beta r) d\beta \tag{15}$$

and substituting these equations into (10) and (11), the following characteristics equation for the $N$-th layer is achieved:

$$s^4 + A_{1,N} s^2 + A_{2,N} = 0 \tag{16}$$

with the following coefficients,

$$A_{1,N} = \left(\frac{\varepsilon_N}{\varepsilon_{\|,N}}\right)\left((k_0^2 \varepsilon_{\|,N}\mu_N - \beta^2) + (k_0^2 \varepsilon_{\perp,N}\mu_N - \beta^2)\frac{\varepsilon_{\|,N}}{\varepsilon_N} + k_0^2 \mu_N \varepsilon_{\|,N}\left(\frac{\varepsilon_{a,N}}{\varepsilon_N}\right)^2\right) \tag{17}$$

$$A_{2,N} = \left(\frac{\varepsilon_N}{\varepsilon_{\|,N}}\right)\left[\left(k_0^2 \varepsilon_{\|,N} \mu_N - \beta^2\right)\cdot\left(k_0^2 \varepsilon_{\perp,N} \mu_N - \beta^2\right)\right] \tag{18}$$

In (14)-(15), $m$ is an integer and $\beta$ is the propagation constant of plasmonic waves. Now, the roots of the characteristics equation are obtained for the $N$-th layer,

$$k_{z,2N-1} = \sqrt{\frac{-A_{1,N} + \sqrt{A_{1,N}^2 - 4A_{2,N}}}{2}} \tag{19}$$

$$k_{z,2N} = \sqrt{\frac{-A_{1,N} - \sqrt{A_{1,N}^2 - 4A_{2,N}}}{2}} \tag{20}$$

Suppose that each gyro-electric layer has the following propagation constants:

$$k_z = \begin{cases} k_{z,1}, k_{z,2} & i=1,2 \; ; \; \text{for layer } N=1 \\ \ldots & \ldots \\ k_{z,2N-1}, k_{z,2N} & i=2N-1, 2N \; ; \; \text{for layer } N \\ k_{z,2N+1}, k_{z,2N+2} & i=2N+1, 2N+2 \; ; \; \text{for layer } N+1 \end{cases} \tag{21}$$

In (21), $N, i$ shows the number of the layer and the index of the root for that layer, respectively. Now, the electromagnetic fields for various regions are written as follows:

$$H_z^m(z) = \begin{cases} A_{m,1,1}^+ e^{+k_{z,1}z} + A_{m,2,1}^+ e^{+k_{z,2}z} & z<0 \\ A_{m,3,2}^+ e^{+k_{z,3}z} + A_{m,4,2}^+ e^{+k_{z,4}z} + \\ A_{m,3,2}^- e^{-k_{z,3}z} + A_{m,4,2}^- e^{-k_{z,4}z} & 0<z<d_2 \\ \ldots \\ A_{m,2N+1,N+1}^- e^{-k_{z,2N+1}z} + \\ A_{m,2N+2,N+1}^- e^{-k_{z,2N+2}z} & z > \sum_{k=2}^{N} d_k \end{cases} \tag{22}$$

$$E_z^m(z) = \begin{cases} T_{1,1}^+ A_{m,1,1}^+ e^{+k_{z,1}z} + T_{2,1}^+ A_{m,2,1}^+ e^{+k_{z,2}z} & z<0 \\ T_{3,2}^+ A_{m,3,2}^+ e^{+k_{z,3}z} + T_{4,2}^+ A_{m,4,2}^+ e^{+k_{z,4}z} + \\ T_{3,2}^- A_{m,3,2}^- e^{-k_{z,3}z} + T_{4,2}^- A_{m,4,2}^- e^{-k_{z,4}z} & 0<z<d_2 \\ \ldots \\ T_{2N+1,N+1}^- A_{m,2N+1,N+1}^- e^{-k_{z,2N+1}z} + \\ T_{2N+2,N+1}^- A_{m,2N+2,N+1}^- e^{-k_{z,2N+2}z} & z > \sum_{k=2}^{N} d_k \end{cases} \tag{23}$$

and

$$T_{i,N}^\pm = \frac{\pm 1}{k_0 \varepsilon_{\|,N} \left(\frac{\varepsilon_{a,N}}{\varepsilon_N}\right)} \left(k_{z,i} + \frac{1}{k_{z,i}}\left(k_0^2 \varepsilon_{\perp,N} \mu_N - \beta^2\right)\right) \quad i=2N-1, 2N, \; N=1,2,3,\ldots \tag{24}$$

The transverse components of electromagnetic fields can be derived:

$$\begin{pmatrix} E_{r,i}^{\pm} \\ H_{r,i}^{\pm} \end{pmatrix} = \bar{\bar{Q}}_{i,N}^{Pos,\pm} \frac{\partial}{\partial r}\begin{pmatrix} E_{z,i}^{\pm} \\ H_{z,i}^{\pm} \end{pmatrix} + \frac{m}{r}\bar{\bar{Q}}_{i,N}^{Neg,\pm}\begin{pmatrix} E_{z,i}^{\pm} \\ H_{z,i}^{\pm} \end{pmatrix} \qquad i = 2N-1, 2N \qquad (25)$$

$$j\begin{pmatrix} E_{\varphi,i}^{\pm} \\ H_{\varphi,i}^{\pm} \end{pmatrix} = \bar{\bar{Q}}_{i,N}^{Neg,\pm} \frac{\partial}{\partial r}\begin{pmatrix} E_{z,i}^{\pm} \\ H_{z,i}^{\pm} \end{pmatrix} + \frac{m}{r}\bar{\bar{Q}}_{i,N}^{Pos,\pm}\begin{pmatrix} E_{z,i}^{\pm} \\ H_{z,i}^{\pm} \end{pmatrix} \qquad i = 2N-1, 2N \qquad (26)$$

Where

$$\bar{\bar{Q}}_{i,N}^{Pos,\pm} = \frac{1}{2}\left[\frac{1}{k_{z,i}^2 + k_0^2 \varepsilon_{+,N}\mu_N}\begin{pmatrix} \pm k_{z,i} & -\omega\mu_0\mu_N \\ \omega\varepsilon_0\varepsilon_{+,N} & \pm k_{z,i} \end{pmatrix} + \frac{1}{k_{z,i}^2 + k_0^2 \varepsilon_{-,N}\mu_N}\begin{pmatrix} \pm k_{z,i} & \omega\mu_0\mu_N \\ -\omega\varepsilon_0\varepsilon_{-,N} & \pm k_{z,i} \end{pmatrix}\right] \qquad (27)$$

$$\bar{\bar{Q}}_{i,N}^{Neg,\pm} = \frac{1}{2}\left[\frac{1}{k_{z,i}^2 + k_0^2 \varepsilon_{+,N}\mu_N}\begin{pmatrix} \pm k_{z,i} & -\omega\mu_0\mu_N \\ \omega\varepsilon_0\varepsilon_{+,N} & \pm k_{z,i} \end{pmatrix} - \frac{1}{k_{z,i}^2 + k_0^2 \varepsilon_{-,N}\mu_N}\begin{pmatrix} \pm k_{z,i} & \omega\mu_0\mu_N \\ -\omega\varepsilon_0\varepsilon_{-,N} & \pm k_{z,i} \end{pmatrix}\right] \qquad (28)$$

are the Q-matrices used in (25)-(26) and,

$$\varepsilon_{\pm,N} = \varepsilon_N \pm \varepsilon_{a,N} \qquad (29)$$

We apply the boundary conditions for the graphene sheets ($N = 1,2,3,...$):

$$E_{r,N} = E_{r,N+1}, \quad E_{\varphi,N} = E_{\varphi,N+1} \qquad (30)$$

$$H_{r,N+1} - H_{r,N} = -\sigma_{H,N}E_{r,N} + \sigma_{O,N}E_{\varphi,N} \qquad (31)$$

$$H_{\varphi,N+1} - H_{\varphi,N} = -\left(\sigma_{O,N}E_{r,N} + \sigma_{H,N}E_{\varphi,N}\right) \qquad (32)$$

Now, the final matrix is obtained:

$$\bar{\bar{S}}_{4N+4,4N+4} \cdot \begin{pmatrix} A_{m,1,1}^{+} \\ A_{m,2,1}^{+} \\ ... \\ A_{m,2N+1,N+1}^{-} \\ A_{m,2N+2,N+1}^{-} \end{pmatrix}_{4N,1} = \begin{pmatrix} 0 \\ 0 \\ ... \\ 0 \\ 0 \end{pmatrix}_{4N,1} \qquad (33)$$

In (33), the matrix $\bar{\bar{S}}$ is

$$\bar{\bar{S}} = \begin{pmatrix} P_{1,1,1}^{+} & P_{2,1,1}^{+} & ... & 0 & 0 \\ R_{1,1,1}^{+} & R_{2,1,1}^{+} & ... & 0 & 0 \\ ... & ... & ... & ... & ... \\ 0 & 0 & ... & ... & ... \\ 0 & 0 & ... & ... & R_{2N+2,N+1,2}^{-} e^{-k_{z,2N+2}\left(\sum_{k=2}^{N} d_k + d_S\right)} \end{pmatrix} \qquad (34)$$

and the following relations have been used in (34),

$$\begin{pmatrix} P_{i,N,1}^{\pm}(r) \\ P_{i,N,2}^{\pm}(r) \end{pmatrix} = \left[\bar{\bar{Q}}_{i,N}^{Pos,\pm}\beta J_m'(\beta r) + \frac{m}{r}\bar{\bar{Q}}_{i,N}^{Neg,\pm}J_m(\beta r)\right]\begin{pmatrix} T_{i,N}^{\pm} \\ 1 \end{pmatrix} \qquad i = 2N-1, 2N \qquad (35)$$

$$\begin{pmatrix} R^{\pm}_{i,N,1}(r) \\ R^{\pm}_{i,N,2}(r) \end{pmatrix} = -j\left[\bar{\bar{Q}}^{Neg,\pm}_{i,N}\beta J'_m(\beta r) + \frac{m}{r}\bar{\bar{Q}}^{Pos,\pm}_{i,N}J_m(\beta r)\right]\begin{pmatrix} T^{\pm}_{i,N} \\ 1 \end{pmatrix} \quad i = 2N-1, 2N \quad N = 1, 2, 3, \ldots \quad (36)$$

Now, the propagation constant and other propagating features such as the effective index can be derived by setting $det(\bar{\bar{S}}) = 0$.

It is worth to be mentioned that the present model gives closed-form mathematical relations for all plasmonic modes propagating inside the waveguide. In all electromagnetic software, the structure should be discretized by numerical methods. Thus, compared to the common software, our model is faster than any numerical method. To be brief, the theoretical expressions presented in this section have two main privileges: first, they can compute the electromagnetic field distributions and dispersion relation for all propagation modes, and second, they have high accuracy and speed compared to any numerical software.

## 3. Validation of the Proposed Model

In the previous section, we proposed an analytical model for general gyro-electric structures, where each graphene sheet was sandwiched between two different gyro-electric layers. Before embarking on presenting analytical results for a special case, the accuracy of the model should be considered. First, we compare our results (for two familiar structures) with the reported results in the literature, second, we simulate two multi-layer structures in COMSOL and LUMERICAL software, respectively, and compare their results with our model.

### 3.1. Comparison with the Published Articles in the Literature

This sub-section aims to check the validity of the presented model (outlined in section 2) by comparing the analytical results against the reported results in the literature. Consider Fig. 2, which illustrates the configuration of a familiar graphene-based waveguide. In this structure, a graphene sheet has been deposited on SiO$_2$-Si layers with the permittivities of 2.09 and 11.9, respectively. We have depicted the effective index ($n_{eff} = Re[\beta]/k_0$) of TM mode for this waveguide in Fig. 2. Furthermore, the chemical potential of graphene is 0.23 eV and the relaxation time is 0.5 ps. The temperate is 300K. The thickness of the SiO$_2$ layer is 5μm. One can see that the effective index increases with the frequency increment. The important point observed in this figure is the good agreement of our analytical model and numerical results prepared by [62], which validates the accuracy of our analytical model.

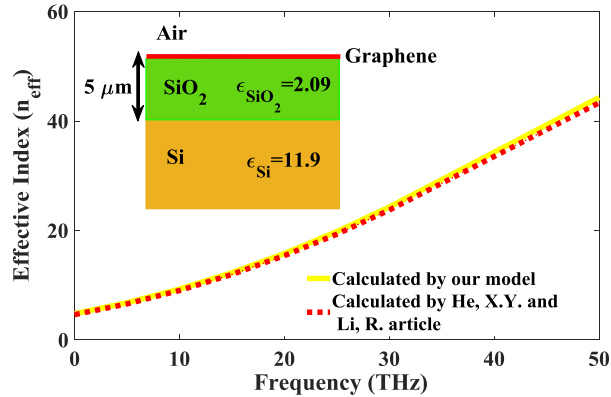

**Fig. 2.** The Comparison of our analytical model and the results of [62] for the effective index of TM mode. The studied case is a familiar waveguide constituting Air-Graphene-SiO$_2$-Si layers.

As the second example, we compare our analytical result with the result of [63]. Fig. 3 represents the configuration of this structure, where an anisotropic graphene sheet has been located on the Silicon layer with the permittivity of 11.9. A magnetostatic bias of 3T is applied perpendicular to its surface. Therefore, the structure supports hybrid TM-TE mode. The temperate is $T = 300K$. The chemical potential of graphene is 0.25 eV and the relaxation time is assumed to be 0.135 ps. It can be observed from Fig. 3 that the effective index increases as the frequency increases. Again, there is an excellent agreement between our analytical result and the result reported by [63], which confirms the validity of the presented model.

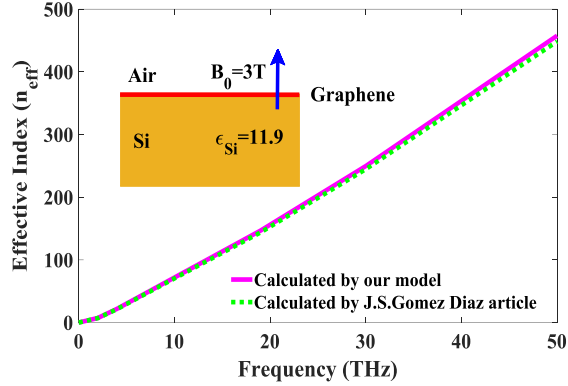

**Fig. 3.** The Comparison of our analytical model and the results of [63] for the effective index of hybrid TM-TE mode. The studied case is a structure where an anisotropic graphene sheet is placed on a silicon layer. The temperate is 300K. The chemical potential of graphene is 0.25 eV and the relaxation time is 0.135 ps. The applied magnetostatic field is 3 T.

### 3.2. Comparison with the Simulation Results

In this sub-section, we study two exemplary structures and compare our theoretical results with simulation results prepared by two numerical software (COMSOL and LUMERICAL). The first structure is shown in Fig. 4, where a dielectric layer (with the permittivity of 3.9) is sandwiched between two SiO$_2$ layers (with the permittivity of 2.09). Two graphene sheets have been located on the border of the middle dielectric and SiO$_2$ layers. In this waveguide, the thickness of graphene sheets is $\Delta = 0.5\ nm$ and the temperature is supposed to be $T = 300\ K$. Moreover, each graphene layer has different chemical potential: $\mu_{c,1} = 0.4\ eV, \mu_{c,2} = 0.45\ eV$.

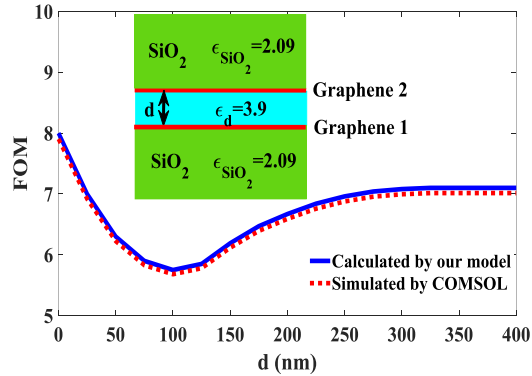

**Fig. 4.** FOM as a function of the dielectric width (d) for a plasmonic waveguide constituting SiO$_2$-Graphene-Dielectric-Graphene-SiO$_2$ layers. The thickness of graphene sheets is $\Delta = 0.5\ nm$. The temperature is $T = 300\ K$. Each graphene layer has different chemical potential: $\mu_{c,1} = 0.4\ eV, \mu_{c,2} = 0.45\ eV$.

The simulation is done by COMSOL software which utilized the finite element method (FEM). The mesh size is not uniform; the assumed mesh size inside the graphene sheets is supposed to be 0.05 nm and its value increases outside the graphene layers (to decrease the computing time and preserve the storage space). As an appropriate boundary condition, the perfectly matched layers (PML) have been located around the structure.

Fig. 4 illustrates the Figure of Merit (defined as $FOM = Re(\beta)/2\pi Im(\beta)$) of the first propagating mode as a function of the middle dielectric width (d). As seen in this figure, there is a minimum value for FOM which occurs near 100 nm. Moreover, there is a good agreement between the simulation and analytical results which confirms the validity of our presented model.

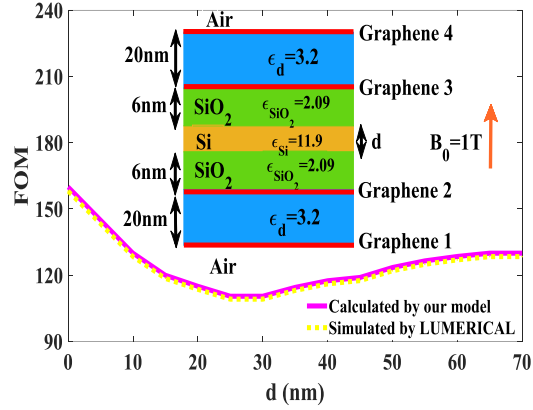

**Fig. 5.** FOM as a function of the Silicon width (d) for a multi-layer structure. The thickness of graphene sheets is $\Delta = 0.5\ nm$. The temperature is $T = 300\ K$ and the magnetic bias is supposed to be 1 T. The carrier mobility of anisotropic graphene layers is $1\ m^2/V.s$. Each graphene layer has different chemical potential: $\mu_{c,1} = 0.75\ eV, \mu_{c,2} = 0.8\ eV,\ \mu_{c,3} = 0.83\ eV, \mu_{c,4} = 0.86\ eV$.

The second example (as shown in Fig. 5) is a test case for a complicated structure (it contains 11 layers). In this figure, the FOM of the hybrid TM-TE mode for a multi-layer structure (constituting Air-Graphene-Dielectric-Graphene-SiO$_2$-Si- SiO$_2$- Graphene-Dielectric-Graphene-Air) as a function of the silicon width (d). The geometrical parameters have been given in this figure. The thickness of graphene sheets is $\Delta = 0.5\ nm$ and the temperature is supposed to be $T = 300\ K$. The dielectric layer has a permittivity of 3.2. The magnetic bias is 1 T. The permittivities of SiO$_2$ and Si layers are 2.09 and 11.9, respectively. The carrier mobility of anisotropic graphene layers is $1\ m^2/V.s$. It is supposed that each graphene layer has different chemical potential: $\mu_{c,1} = 0.75\ eV, \mu_{c,2} = 0.8\ eV,\ \mu_{c,3} = 0.83\ eV, \mu_{c,4} = 0.86\ eV$.

This structure is simulated by LUMERICAL FDTD Solutions. The utilized numerical method in this simulation is finite-difference time-domain (FDTD). Similar to the structure of Fig. 4, the mesh size is not uniform and its size is supposed to be 0.05 nm inside the graphene sheets. An excellent agreement between our analytical result and the simulation result (prepared by LUMERICAL software) can be observed from Fig. 5, which shows the high accuracy of the model.

## 4. Results and Discussions

In the previous section, we checked the validity of the proposed model. Here, we will investigate the analytical results for a gyro-electric structure, as a special case of the general waveguide. It is obvious that the study of all possible structures is impossible and is outside the scope of this article. Therefore, we apply the analytical model for a specific graphene-based gyro-electric waveguide.

Consider the gyro-electric graphene-based waveguide (as a special case of the general structure of Fig. 1), as demonstrated in Fig. 6. In this configuration, the magnetic bias is applied in the z-direction. The magnetized gyro-electric is supposed to be the n-type InSb, with a thickness of $s = 50\ nm$ and parameters of $\varepsilon_\infty = 15.68$, $m^* = 0.022 m_e, n_s = 1.07 \times 10^{17}/cm^3, \nu = 0.314 \times 10^{13} s^{-1}$ where $m_e$ is the electron's mass. The thickness of the SiO$_2$ layer is $t = 200\ nm$. In this waveguide, graphene has the following parameters: $\mu_c = 0.45\ eV,\ \tau = 0.34\ ps,\ T = 300\ K$.

In Fig. 7, the modal properties of the gyro-electric graphene-based waveguide have been depicted as a function of the frequency for the magnetic bias of 4T. To consider the effect of non-reciprocity, the propagation features also have been calculated for the reverse direction (i.e. $B_0 = -4T$). As observed in this figure, the proposed structure of Fig. 6 is nonreciprocal due to the change of plasmonic properties by reversing the direction of the external magnetic bias. It is evident from Fig. 7(a) that the nonreciprocity occurs in the vicinity of 6.5 THz, where propagation loss has a maximum point at this frequency (see Fig. 7 (b)).

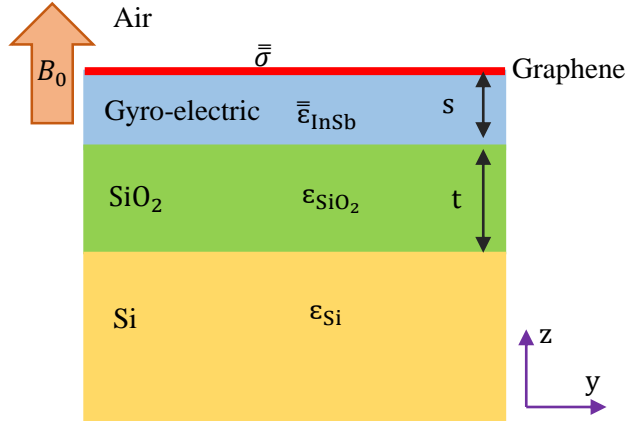

**Fig. 6.** The cross-section of the gyro-electric graphene-based waveguide, constituting Graphene-InSb-SiO$_2$-Si layers.

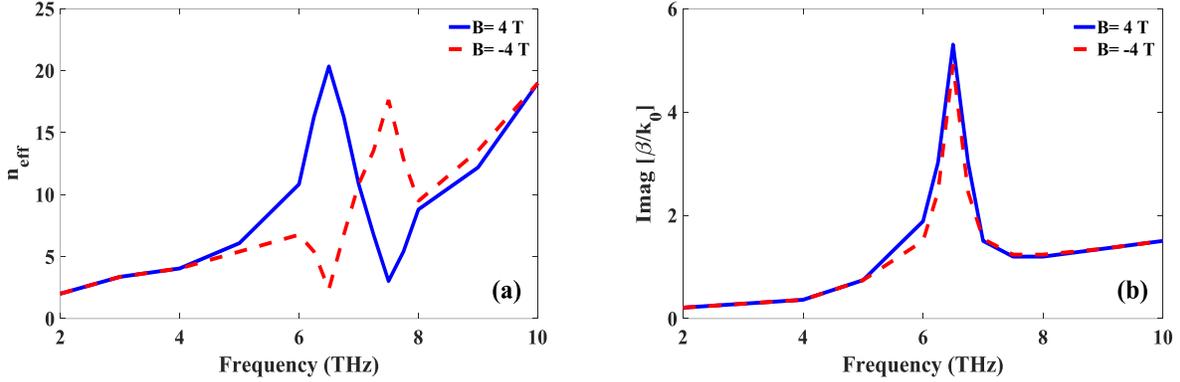

**Fig. 7.** Analytical results for (a) the effective index and (b) the propagation losses versus frequency for various external magnetic fields (B= -4, 4 T). The chemical potential of the graphene layer is supposed to be $\mu_c = 0.45 \, eV$.

To investigate the performance of the plasmonic waveguides, one of the main parameters introduced in the literature is Figure of Merit (defined as $FOM = Re(\beta)/2\pi Im(\beta)$) [66]. In Fig. 8, FOM has been shown as a function of external magnetic bias at various frequencies. It should be noted that our studied frequency window, in which nonreciprocity occurs, is 2-10 THz (see Fig. 7). However, in Fig. 8, FOM also has been calculated at a higher frequency ($f = 15 \, THz$) to exactly consider the performance of the structure at the outside of this window. It is obvious from this figure that near resonance frequency (6.5 and 7 THz), FOM diagram depends on the value of the external magnetic field. In these frequencies, there are valleys (or minimum points) for the external magnetic bias, which FOM reaches its minimum value.

The strong variations of FOM on external magnetic field near non-reciprocity frequency range originates from the main role of non-diagonal elements in permittivity tensor of the gyro-electric layer at the resonance frequency. As the frequency increases and it moves away from resonance frequency (see Fig. 8, the analytical result of FOM at the frequency of 15 THz), the magnitude of non-diagonal elements in permittivity tensor becomes smaller compared to diagonal elements and thus the FOM does not depend strongly on the magnetic bias at the frequency of 15 THz.

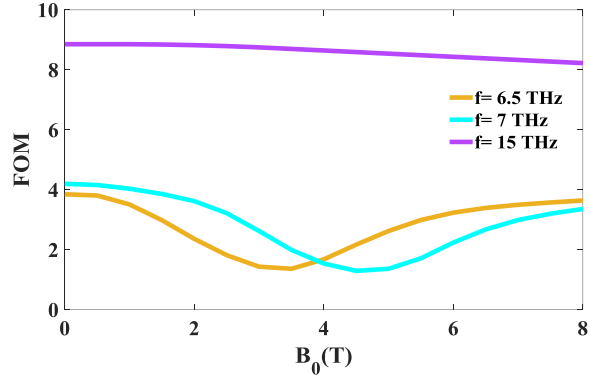

**Fig. 8.** FOM as a function of magnetic bias at various frequencies. The chemical potential of the graphene layer is supposed to be $\mu_c = 0.45\ eV$.

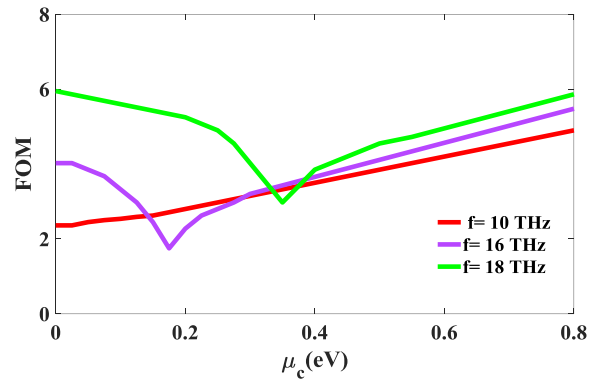

**Fig. 9.** FOM as a function of chemical potential at various frequencies. The applied magnetics bias is *B= 4 T*.

The tunability of the proposed structure via the chemical potential of graphene is demonstrated in Fig. 9, which represents the variations of FOM as a function of the chemical potential at various frequencies. At lower frequencies, i.e. frequency range of non-reciprocity effect (2-10 THz), FOM increases as the chemical potential increases (see Fig. 9, the analytical result of FOM at the frequency of 10 THz). As frequencies increases and keeps out of the non-reciprocity range, the FOM strongly depends on chemical potential. For instance, consider the frequency of 16 THz in this figure. This diagram has a minimum point in the vicinity of 0.18 eV and thus the structure should not be designed at this chemical potential.

As a final point, we perform a parametric study to investigate the influence of InSb and Silica thicknesses on FOM. Tables 1, 2 report FOM for various values of InSb and silica thicknesses, respectively (at the frequencies 6.5 and 10 THz). One can see from Table. 1 that as the thickness of the gyroelectric layer increases, FOM decreases. It happens because the field confinement inside the InSb decreases by the thickness increment. However, the slope of FOM variations is different at various frequencies. For instance, at 6.5 THz, the slope of changes is greater compared to 10 THz. The main reason behind this is the high oscillation of SPPs near resonance frequency. It is evident from Table. 2 that the increment of silica thickness will decrease FOM. Furthermore, FOM variations are negligible here, compared to Table 1. This matter is concerned with the energy concentration inside the gyroelectric layer, especially on the graphene sheet. Hence, the change of InSb thickness will alter FOM more in comparison to variations of silica thickness.

Table 1. FOM for Various Thicknesses of Gyroelectric Layer (InSb) @ $\mu_C = 0.2\ ev, t = 200\ nm$

| S (nm) | | 50 | 60 | 70 | 80 | 90 | 100 |
|---|---|---|---|---|---|---|---|
| **FOM** | @ 6.5 THz | 4.10 | 3.91 | 3.72 | 3.45 | 3.34 | 3.09 |
| | @ 10 THz | 8.50 | 8.43 | 8.35 | 8.29 | 8.19 | 8.05 |

Table 2. FOM for Various Thicknesses of Silica @ $\mu_C = 0.2\ ev, S = 50\ nm$

| t (nm) | | 200 | 210 | 220 | 230 | 240 | 250 |
|---|---|---|---|---|---|---|---|
| **FOM** | @ 6.5 THz | 4.10 | 4.08 | 4.06 | 4.00 | 3.93 | 3.84 |
| | @ 10 THz | 8.50 | 8.49 | 8.48 | 8.46 | 8.42 | 8.36 |

## 5. Conclusion

In this article, we proposed a novel analytical model for gyro-electric multi-layer graphene-based structures. In our general waveguide, each graphene sheet was located between two different magnetized gyro-electric layers. The validity of our model was checked by the reported results of published articles in the literature and also by two numerical solvers. To demonstrate the richness of the general structure, a new gyro-electric graphene-based waveguide, constituting Graphene-InSb-SiO$_2$-Si layers, was studied. It was shown that the proposed gyro-electric waveguide is non-reciprocal, i.e. its propagation properties vary as the direction of external magnetic bias reverses. FOM, an important parameter to evaluate the performance of the structure, was depicted as a function of magnetic bias and chemical potential. The FOM was strongly dependent on the external magnetic bias at the cyclotron frequency. Furthermore, it was observed that the gyro-electric graphene-based waveguide is a tunable device and its FOM can be adjusted via the chemical potential of graphene. Harnessing gyro-electric media together with graphene sheets can be utilized for the design of innovative plasmonic devices at the THz frequencies.